\documentclass[preprint,12pt]{elsarticle}
\usepackage{amssymb}
\usepackage{amsmath}
\usepackage{amsthm}
\usepackage{amsfonts} 
\usepackage{bm}
\usepackage{physics}
\usepackage{graphics}
\usepackage{braket} 
\usepackage{graphicx}
\usepackage{dcolumn}
\usepackage{multirow}
\usepackage{appendix}
\journal{Physics Letters A}

\usepackage{xcolor}
\DeclareRobustCommand{\rchi}{{\mathpalette\irchi\relax}}
\newcommand{\irchi}[2]{\raisebox{\depth}{$#1\chi$}} 

\begin{document}
\begin{frontmatter}
\title{Linear Response Conductance of Metallic Single Electron Pump}
\tnotetext[]{}
\author{P. Srivilai*\textsuperscript{1}}
\ead{prathan.s@msu.ac.th}
\author{T. Thongsuk\textsuperscript{2}}
\ead{tawan.th@ksu.ac.th}
\author{P. Harata\textsuperscript{1}}
\ead{pipat.harata.c@gmail.com}
\address{\textsuperscript{1}NanoMaterials Physics Research Unit (NMPRU), Department of Physics,
Faculty of Science, Mahasarakham University, Khamriang Sub-District, Kantarawichai District, Mahasarakham 44150, Thailand}
\address{\textsuperscript{2}Department of General Science, Faculty of Education and Educational Innovation, Kalasin University, Thailand}
\begin{abstract}

We calculate the linear-response conductance of a metallic single-electron pump using the path-integral Monte Carlo (PIMC) method. The Coulomb oscillations of the conductance are calculated to illustrate the influence of the Coulomb blockade effect on the system. Furthermore, the experimental conductance is compared with the calculated conductance of various gate voltage configurations and temperatures. The results are consistent even in the low-temperature regime, where significant quantum fluctuation occurs, and the semiclassical approximation fails. Consequently, the present investigation replicates the success of the PIMC approach while precisely describing the quantum fluctuation phenomena of single-electron devices.

\end{abstract}

\end{frontmatter}

\section{Introduction}
\label{sec:intro}

A single-electron pump (SEP) mainly consists of two islands (quantum dots), three tunnel junctions, and voltage electrodes to control the shuttle of the individual electrons through the system \cite{Grabert, Howe2021}. They can potentially revolutionize fields like quantum metrology as they can precisely measure current at the fundamental level and quantum information processing \cite{Milton_2010, RevModPhys.85.1421, Giblin_2012, Harrington2022, DeMille2024}, enabling the creation and manipulation of qubits \cite{Pashkin2003, Andreev2004, Hines_2007, Shuo2011, Utsugi_2023, Blumenthal1a2023}, and these devices have been essential in advancing our understanding of electron behaviour at the quantum level. SEPs have been widely studied both theoretically and experimentally \cite{Pothier1991, Pothier1992, Limbach2005, Kaestner2008, Prada2012, Tanttu2015, Joost2017, Schoinas2024}. One of the most frequent experiments is the Coulomb blockade oscillation, where current or conductance is measured via bias voltage changes to investigate the influence of the Coulomb blockade effect \cite{Grabert, PhysRevB.60.16906, Fujisawa_2006,Bäuerle_2018, PhysRevLett.121.257701, Shang2013, Banszerus2020, PhysRevApplied.20.044043}. 

Limbach and coworkers had studied the linear response conductance of the metallic SEP, \cite{Limbach2005, Limbach2002}, and only the lowest order perturbation theory, the so-called sequential model \cite{Grabert}, was verified to describe the experiment's results quantitatively. However, quantum fluctuations due to the coupling of charge states, which are fundamental constituents of the devices, were ignored in the description. Nevertheless, the experiment's Coulomb oscillation of conductance verified the single electron transistor (SET) theory in great detail \cite{Wallisser2002,Christoph2004}. It is still unknown, though, if the theoretical calculation will hold up to more complex experiments than that of the SET, like SEP. Therefore, a demonstration is necessary for additional theoretical study in the nanoscience field. Since all parameters are experimentally accessible, and Limbach's experiment enables us to compare the experimental and theoretical results over experimentally accessible parameters, which is the aim of this Letter.

The Letter's structure is as follows: the SEP's Hamiltonian and all experimental parameters necessary for the conductance calculation are introduced in Sec.\ref{sec:model}. The imaginary time current correlation function is represented as a functional integral over phase fields in Sec.\ref{sec:CAF} by extending the partition function of the SEP. Sec.\ref{sec:MCS} and Sec.\ref{sec:LRC} present the current correlation function suitable for the PIMC simulation and the relationship of the DC conductance and the correlation function, respectively. The theoretical results are compared with the experiment data and discussed in Sec.\ref{sec:results}. Finally, we conclude and discuss possible extensions in Sec.\ref{sec:conclusion}.

\section{  Model Hamiltonian and Experimental Parameters }
\label{sec:model}

\begin{figure}[h]
\centering
\includegraphics[width=0.95\textwidth]{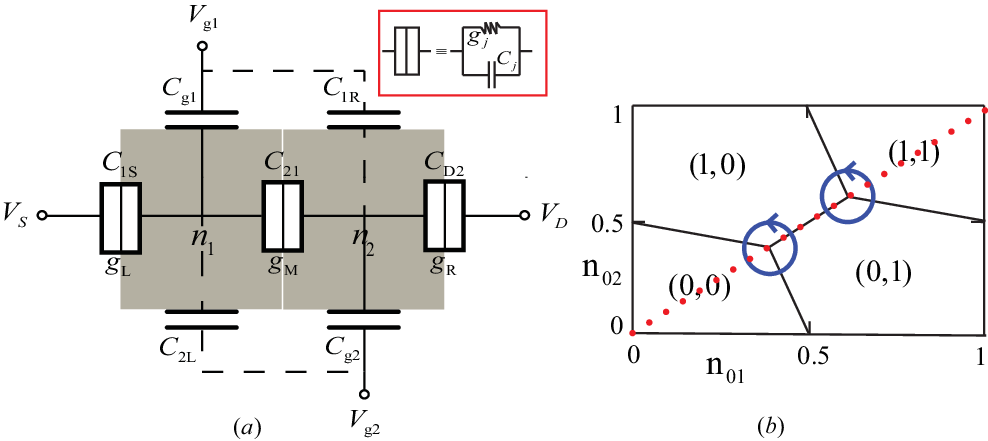}
\vspace{0.2cm}
\caption{(a) Equivalent circuit diagram of SEP system consists of two gate voltages coupled directly to the first and second islands by the capacitance $C_{g1}$ and $C_{g2}$ and the experimentally unavoidable stray capacitor $C_{2L}$ and $C_{1R}$, respectively. (b) The red dotted line in the stability diagram of the SEP passes two classical triple points where three adjacent states can be occupied. $(n_1, n_2)$ denotes a charge ground state in which the first and second islands have the excess electron numbers $n_1 $ and $n_2 $, respectively. }
\label{fig:SEP}
\end{figure}

The circuit diagram in Fig.\ref{fig:SEP}\,(a) illustrates the SEP system consisting of three tunnelling junctions and two islands. This arrangement has the differential voltage denoted by $V_{DS}= V_{D}-V_{S}$. Two voltage gates can tune the electrostatic potentials on the two islands. Consequently, $n_{1}$ and $n_{2}$ indicate that the excess electron numbers on the first and second islands, respectively, can be adjusted. The quantum transport characteristics of the SEP can be explained by the electrostatic charge and electron tunnelling between the leads and islands. Since the SEP comprises metallic islands with spacing between the island states significantly less than the charging energy, energy quantization can be neglected. We can then microscopically describe the electronic motion in the metallic SEP by Hamiltonian as \cite{Grabert}

\begin{equation} \label{Hamiltonian}
H=H_{B}+H_{T} + H_{C}\, .
\end{equation}
The term $H_{B}$ is the Hamiltonian for the non-interacting electrons in the leads and the islands
\begin{equation}
\label{HsubB} H_{B}=\sum_{Jk\sigma}\epsilon_{Jk\sigma}c_{Jk\sigma}^{\dag}c_{Jk\sigma}^{}+
\sum_{Ik\sigma}\epsilon_{Ik\sigma}d_{Ik\sigma}^{\dag}d_{Ik\sigma}^{}\, .
\end{equation}
Here $\epsilon_{Jk\sigma}$ is the energy of an electron with longitudinal wave vector $k$ in channel $\sigma$ of lead $J$, where $J \in \{S,D\}$ . The channel index $\sigma$ includes the transversal and spin quantum numbers. The notation $c_{Jk\sigma}^{\dag}$ $(c_{Jk\sigma})$ is a creation (annihilation) operator for an electron of the continuous state $\ket{k,\sigma}$ in the lead $J$. Likewise, $\epsilon_{Ik\sigma}$ is the energy of an electron with the longitudinal wave vector $k$ in channel $\sigma$ of island $I$, where $I \in \{1,2\}$, and $d_{Ik\sigma}^{\dag}$ $ (d_{Ik\sigma})$ is the creation (annihilation) operator for an electron of the continuous state $\ket{k,\sigma}$ in the island $I$.

The second term describes the electron tunnelling across the three tunnelling junctions in the system as shown in Fig.\ref{fig:SEP},
\begin{eqnarray}
\label{Htunneling}
H_{T}=\sum_{kq\sigma}&\Big[&d_{1k\sigma}^{\dag}t_{1Skq \sigma}{\rm e}^{-i\varphi_{1}}c_{Sq\sigma}^{}
 +d_{2k\sigma}^{\dag}t_{21kq \sigma}{\rm e}^{-i(\varphi_{2}-\varphi_{1})}d_{1q\sigma}^{}\nonumber \\ 
 &&+c_{Dk\sigma}^{\dag}t_{D2kq \sigma}{\rm e}^{i\varphi_{2}}d_{2q\sigma}^{}+\hbox{H.c.}\label{HsubT}\Big],
\end{eqnarray}
where $t_{IJjkq\sigma}$ is the tunnelling amplitude of the tunnelling junction between conductors $I$ and $J$ for an electron tunnelling between states $|q\sigma\rangle$ and $|k\sigma\rangle$. The phase variable $\varphi_{I}$ is the conjugate operator to the number operator $n_{I}$ of the excess charge on the island $I$, with the canonical commutation relation $[n_{I},\varphi_{I}]=i$. Accordingly, the charge shift operator ${\rm e}^{-i\varphi_{I}}$ adds one charge to the island $I$. Here, we define  $\varphi_S=\varphi_D=0$ and use $\hbar = 1$ throughout this Letter.

The last term in the Hamiltonian is the Coulomb charging energy expressed as \cite{Fujisawa2002, Wallisser2002}
\begin{eqnarray}\label{Ec0}
   H_{C}&=&{{E_{11}}{( {n_1}-{n_{01}})}^{2}}+{{E_{22}}{( {n_2}-{n_{02}})}^{2}}\nonumber\\
   & &+2{{E_{21}}( {n_1}-{n_{01}})( {n_{2}}-{n_{02}})},
\end{eqnarray}
where $n_I$ is the (excess) electron number on the island $I$. The induced charges on the first and second islands are defined by $n_{01}= (C_{g1}V_{g1}+C_{2L}V_{g2})/e$, and $n_{02}= (C_{g2}V_{g2}+C_{1R}V_{g1})/e$, respectively. The coefficients $E_{I I^{\prime}}$ are the matrix elements of the matrix $\mathbb{E}_{C}$ defined by
\begin{equation}\label{ECN}
 \mathbb{E}_{C}\equiv
  \begin{pmatrix}
  E_{11} & E_{12} \\
  E_{21} & E_{22}
\end{pmatrix}
=
  \frac{{{e}^{2}}}{2(C_{\Sigma 1}C_{\Sigma 2}-C_{21}^{2})}\begin{pmatrix}
  C_{\Sigma 2} & C_{21} \\
  C_{21} & C_{\Sigma 1}
\end{pmatrix}.
\end{equation}
where $C_{\Sigma 1}=C_{1S}+C_{21}+C_{g1}+C_{2L}$ and $C_{\Sigma 2}=C_{21}+C_{D2}+C_{g2}+C_{1R}$.

To calculate the SEP's linear response conductance using the Kubo formula \cite{Kubo_1966, Goppert1999} and functional integral \cite{Negele1987, Goppert2000}, we express the grand canonical partition function of the SEP, $ Z={\rm tr}\, \{{\rm e}^{-\beta (\hat{H}-\mu\hat{N})}\}$, as a path integral over phase fields conjugated to the island charges, with a path probability given in terms of an analytical action functional. However, as explored in our previous work \cite{Harata_2022}, we will skip the details of calculating the system's partition function here. Instead, in the following sections, we will discuss how the partition function is extended to a generating function for the current correlation function relating to the conductance of the SEP and calculate the correlation function using PIMC simulation.

To compare the theoretical results with the experimental results, we first calculate the Coulomb oscillations of the conductance under the dimensionless gate voltage condition plotted by the red dot in the stability diagram in Fig.\ref{fig:SEP} (b). The detail of determining this condition is shown in \ref{sec:SD}. Under the biased gate condition, the SEP operates with electrons transferring past two classical triple points indicated by blue cycles in Fig.\ref{fig:SEP} (b). An electron can tunnel through the two islands when slightly $V_{DS}$ is present. For example, an electron that can transfer from the source to the drain is represented by the state sequence $(0,0)\rightarrow (1,0) \rightarrow(0,1)$, which is equivalent to circling the classical triple point counter-clockwise. Since the theoretical results will be compared with the experimental results of the two-island system, called sample 2 in Ref.\cite{Limbach2005, Limbach2002}, this section then closes by introducing all the experimental parameters needed to calculate the conductance of the SEP, which are included in Table \ref{tab:table3}.

\begin{table*}
\caption{\label{tab:table3} All parameters in this table are illustrated in Fig.\ref{fig:SEP}(a). The notation $g_{j}$ stands for the dimensionless conductance of the individualtunnelling junction defined as $g_{j}=G_{j}/G_{K}$, where $j \in \{1S,21,D2\}$ and $G_{K} = e^2/h$. $ G_{0}$ is the high-temperature conductance of the SEP \,\cite{Limbach2005}.}

\label{parameters}
\begin{tabular}{ |c|c|c|c|c|c|c|c|c|c|c|c|} 
\hline
Parameters& $C_{1S}$ & $C_{21}$ & $C_{D2}$& $C_{g1}$ & $C_{1R}$ & $C_{2L}$ & $C_{g2}$ & $g_{L}$ & $g_{M}$ & $g_{R} $ & $G_{0} $ \\
\hline
Values& 181 & 173 & 236 & 50.5  & 18.0 & 21.5  & 58.6 & 0.52 & 1.32 & 0.83 & 10.0  \\
\hline
Units& (aF) & (aF) & (aF) & (aF)  & (aF) & (aF)  & (aF) & - & - & - & ($\mu S$) \\
\hline
\end{tabular}
\end{table*}

\section{Current Autocorrelation Function}
\label{sec:CAF}
This section describes how an appropriate source term is introduced to extend the partition function to a generating function for the current correlation function calculation. The current correlation function in the imaginary time between the current in junction $K$ at time $\tau$ and junction $K^{\prime}$ at time $\tau^{\prime}$ can be expressed as a second-order functional derivative of the generating functional as \cite{PhysRevB.58.R10155}

\begin{equation}\label{correlation}
 \langle I_K(\tau)I_{K^{\prime}}(\tau^{\prime})\rangle =
 \frac{1}{Z_{\rm gen}[0]}\frac{\delta^{2}Z_{\rm
 gen}[\rchi]}{\delta\rchi_K(\tau)\delta\rchi_{K^{\prime}}(\tau^{\prime})}\Bigg|_{\rchi\equiv 0} \,,
\end{equation}
where the junction $K\in\{L,M,R\}$ and $Z_{\rm gen}[0]$ is the generating functional without the source terms. The generating functional $Z_{\rm gen}[\rchi]$ is described in more detail in \ref{sec:GF_SEP}. Since the dependence on the source fields $\rchi_K(\tau)$ arises only from the tunnelling action expressed in Eq.\,(\ref{SsubTun4}), we obtain 
\begin{equation}\label{correlation2}
\langle I_K(\tau)I_{K^{\prime}}(\tau^{\prime})\rangle = \left\langle
{\rm e}^{-2\pi i\,(\vec{n}_{g}\cdot\vec{k})}\,\mathcal{C}_{K,\tau;K^{\prime},\tau^{\prime}}[\vec{\varphi}] \right\rangle\, ,
\end{equation}
where the vector $\vec{n}=(n_{01},n_{02} )$, $\vec{k}=(k_{1},k_{2} )^{T}$, and
\begin{eqnarray}
\label{C2ofPhi}
\mathcal{C}_{K,\tau;K^{\prime},\tau^{\prime}}[\vec{\varphi}] &=&   \Bigg(  \frac{\delta S_{\rm tun}[\vec{\varphi},\rchi]}{\delta
\rchi_K(\tau)}
\frac{\delta S_{\rm tun}[\vec{\varphi},\rchi]}{\delta \rchi_{K^{\prime}}(\tau^{\prime})}-\frac{\delta^2 S_{\rm tun}[\vec{\varphi},\rchi]}{\delta \rchi_K(\tau)\delta \rchi_{K^{\prime}}(\tau^{\prime})}\Bigg)_{\rchi\equiv
0}\, .
\end{eqnarray}
Here, we introduced the path average of a phase functional $F[\vec{\varphi}]$ expressed as
\begin{eqnarray}\label{average}
\left\langle\,  F[\vec{\varphi}]\, \right\rangle = \frac{1}{Z_{\rm eff}}
\sum_{k_{1},k_{2}=-\infty}^{\infty}\int\limits_{\varphi_{1}(0)}^{\varphi_{1}(0)+2\pi
k_{1}}\!\!\!\!\!\!\!\!  D[\varphi_{1}]\!\!\int\limits_{\varphi_{2}(0)}^{\varphi_{2}(0)+2\pi k_{2}}\!\!\!\!
\!\!\!\!D[\varphi_{2}]\, {\rm e}^{-S^{}_{\rm eff}[\vec{\varphi}]}\, F[\vec{\varphi}] \, ,
\end{eqnarray}
where
\begin{equation}\label{Zeff}
Z_{\rm eff}=\sum_{k_{1},k_{2}=-\infty}^{\infty}\!\!\!\!
\int\limits_{\varphi_{1}(0)}^{\varphi_{1}(0)+2\pi k_{1}}\!\!\!\! \!\!\!\! D[\varphi_{1}]\!\!
\int\limits_{\varphi_{2}(0)}^{\varphi_{2}(0)+2\pi k_{2}}\!\!\!\!\!\!\!\! D[\varphi_{2}]\,
{\rm e}^{-S^{}_{\rm eff}[\vec{\varphi}]}\, ,
\end{equation}
and the effective action  
\begin{equation}\label{effactionphi}
S_{\rm eff}[\vec{\varphi}]=S_C[\vec{\varphi}]+ S^{}_{\rm tun}[\vec{\varphi}]\, ,
\end{equation}
with the Coulomb and tunnelling actions are expressed in Eq.\,(\ref{SC2}) and Eq.\,(\ref{STun0}), respectively. From the tunnelling action in Eqs.\,(\ref{STun1})--\,(\ref{STun2}), we obtain
\begin{eqnarray}
\label{VarSTun1}
\mathfrak{F}_{K,\tau}[\vec{\varphi}] &\equiv&\frac{\delta S^{(1)}_{\rm tun}[\vec{\varphi},\rchi]}{\delta \rchi_K(\tau)}\\ &=&
2eg^{}_K\int_0^{\beta}d\tau^{\prime}\,\alpha(\tau-\tau^{\prime}) \sin\left[\psi_K(\tau)-\psi_K(\tau^{\prime})\right]\nonumber \, ,
\end{eqnarray}
where $\psi_L(\tau) = \varphi_1(\tau)$, $\psi_M(\tau) = \varphi_2(\tau)-\varphi_1(\tau)$, and $\psi_R(\tau) = -\varphi_2(\tau)$.  Likewise,
\begin{eqnarray} \label{VarSTun2}
\mathfrak{F}_{K,\tau;K^{\prime},\tau^{\prime}}[\vec{\varphi}] &\equiv&   \frac{\delta^2 S^{(2)}_{\rm tun}[\vec{\varphi},\rchi]}{\delta
\rchi_K(\tau)\delta \rchi_{K^{\prime}}(\tau^{\prime})}\\
&=& - 2e^2\delta_{K,K^{\prime}}g_K\alpha(\tau-\tau^{\prime})
\cos\left[\psi_K(\tau)-\psi_K(\tau^{\prime})\right]\,.\nonumber
\end{eqnarray}
These equations combine with Eq.\,(\ref{C2ofPhi}) to yield
\begin{equation}\label{C2ofPhi2}
\mathcal{C}_{K,\tau;K^{\prime},\tau^{\prime}}[\vec{\varphi}] =
\mathfrak{F}_{K,\tau}[\vec{\varphi}]\mathfrak{F}_{K^{\prime},\tau^{\prime}}[\vec{\varphi}]
- \mathfrak{F}_{K,\tau;K^{\prime},\tau^{\prime}}[\vec{\varphi}] \, ,
\end{equation}
where the first and second terms in the right-hand side in Eq.\,(\ref{C2ofPhi2}) come from the action's first and second-order variational derivatives. 
The path average in Eq.\,(\ref{correlation2}) can thus be symmetrized concerning path inversion, which gives
\begin{equation}\label{correlation3}
\langle I_K(\tau)I_{K^{\prime}}(\tau^{\prime})\rangle= \left\langle
\cos\left(2\pi(\vec{n}_{g}\cdot\vec{k})\right)\,  \mathcal{C}_{K,\tau;K^{\prime},\tau^{\prime}}[\vec{\varphi}] \right\rangle\, .
\end{equation}

Let us consider a path $\vec{\varphi}(\tau)$ with winding numbers $\vec{k}$, i.e.,
\begin{equation}\label{cplr}
\varphi_1(\beta)=\varphi_1(0)+2\pi k_1\,\quad {\rm and}\quad \varphi_2(\beta)=\varphi_2(0)+2\pi k_2\, .
\end{equation}
Using the relationship $\varphi_I(\beta+\tau)=\varphi_I(\tau)+2\pi k_I$, a shifted path can be mapped back to
the time interval ($0,\beta$). Since the path average in Eq.\,(\ref{average}) includes a summation over all shifted paths with the same action, the
correlation function in Eq.\,(\ref{correlation3}) depends only on $\tau-\tau^{\prime}$, i.e.,
\begin{equation}\label{correlator}
\langle I_K(\tau)I_{K^{\prime}}(\tau^{\prime})\rangle = \mathbf{C}_{K,K^{\prime}}(\tau-\tau^{\prime}) \,.
\end{equation}

In the theoretical conductance calculation, the quantity of interest is the tunnelling current flowing from the source to the drain as a function of the applied voltages. Under stationary conditions where displacement currents are absent, the tunnelling current coincides with the tunnelling current through any of the three tunnel junctions of the SEP. Let us introduce the dimensionless series conductance $g$ of the three tunnel junctions defined by
\begin{equation}\label{seriesg}
    \frac{1}{g}=\frac{1}{g_L}+\frac{1}{g_M}+\frac{1}{g_R} .
\end{equation}
A convenient choice of the SEP's correlation function is based on an average of the tunnelling currents defined by
\begin{equation}\label{AvCurrent}
\bar I(\tau) = \sum_{K\in\{L,M,R\}}\frac{g}{g_K} I_K(\tau)
\end{equation}
and the corresponding correlation function
\begin{equation}\label{AvCorrelation}
\mathbf{\bar C}(\tau-\tau^{\prime})=\langle \bar I(\tau)\bar I(\tau^{\prime})\rangle\, .
\end{equation}
Using Eqs.\,(\ref{AvCurrent}) and (\ref{correlator}), the average of the correlation function can be split into two parts as
\begin{eqnarray}
\label{AvCorrelation2spilt}
\mathbf{\bar C}(\tau-\tau^{\prime}) &\equiv& \mathbf{\bar C}^{1^{st}}(\tau-\tau^{\prime}) + \mathbf{\bar C}^{2^{nd}}(\tau-\tau^{\prime})\\
&=&\left\langle\cos\left(2\pi\vec{n}_{g} \cdot\vec{k}\right)\,\mathfrak{F}_{\tau}[\vec{\varphi}] \,\mathfrak{F}_{\tau^{\prime}}[\vec{\varphi}] \right\rangle + \left\langle\cos\left(2\pi\vec{n}_{g} \cdot\vec{k}\right)\,\mathfrak{F}_{\tau;\tau^{\prime}}[\vec{\varphi}] \right\rangle\, \nonumber,
\end{eqnarray}
where
\begin{eqnarray}
\label{Fphi}
\mathfrak{F}_{\tau}[\vec{\varphi}] &= &\sum_{K\in\{L,M,R\}}\frac{g}{g_K}\mathfrak{F}_{K,\tau}[\vec{\varphi}]\\
&=&2eg\int_0^{\beta}d\tau^{\prime}\,\alpha(\tau-\tau^{\prime})\Big(\,
\sin\left[\varphi_1(\tau)-\varphi_1(\tau^{\prime})\right]-\sin\left[\varphi_2(\tau)-\varphi_2(\tau^{\prime})\right]\nonumber\\ &&\
+ \sin\left[\varphi_2(\tau)-\varphi_1(\tau)-\varphi_2(\tau^{\prime})+\varphi_1(\tau^{\prime})\right]\Big)\, ,\nonumber
\end{eqnarray}
and
\begin{eqnarray} \label{FFphi}
\mathfrak{F}_{\tau;\tau^{\prime}}[\vec{\varphi}] &=&\sum_{K,K^{\prime}\in\{L,M,R\}}\frac{g^2}{g_K^2}\mathfrak{F}_{K,\tau;K^{\prime},\tau^{\prime}}[\vec{\varphi}]\nonumber\\
&=& 2e^2g\,\alpha(\tau-\tau^{\prime})\Bigg(\frac{g}{g_L}\cos\left[\varphi_1(\tau)-\varphi_1(\tau^{\prime})\right]
+\frac{g}{g_R}\cos\left[\varphi_2(\tau)-\varphi_2(\tau^{\prime})\right]\nonumber\\ &&\ +
\frac{g}{g_M}\cos\left[\varphi_2(\tau)-\varphi_1(\tau)- \varphi_2(\tau^{\prime})+\varphi_1(\tau^{\prime})\right]\Bigg)\,
\nonumber\\
&\equiv& 4 \pi G_{0}\alpha(\tau-\tau^{\prime})A_{\tau;\tau^{\prime}}[\vec{\varphi}]\, ,
\end{eqnarray}
with the autocorrelation function
\begin{eqnarray}\label{autocf}
A_{\tau;\tau^{\prime}}[\vec{\varphi}] &=&\frac{g}{g_L}\cos\left[\varphi_1(\tau)-\varphi_1(\tau^{\prime})\right]
+\frac{g}{g_R}\cos\left[\varphi_2(\tau)-\varphi_2(\tau^{\prime})\right]\nonumber\\ &&\ +
\frac{g}{g_M}\cos\left[\varphi_2(\tau)-\varphi_1(\tau)- \varphi_2(\tau^{\prime})+\varphi_1(\tau^{\prime})\right]\, ,
\end{eqnarray}
which is a relevant quantity of the PIMC simulation. In addition, after we meticulously calculated the correlation function in Eq.\,(\ref{AvCorrelation2spilt}) by the PIMC simulation in the hold range of the interested parameters, the results show that the first term $\mathbf{\bar C}^{1^{st}}$ always equals zero \cite{ Prathan2012}. Therefore, the correlation function and its consequences presented in this Letter were obtained by calculating only the second term in Eq.\,(\ref{AvCorrelation2spilt}).

\section { Suitable Correlation Function for PIMC Simulation }
\label{sec:MCS}
For convenience in the PIMC simulation, we measured all energies in units of the charging energy $E_C$ of the SEP defined by \cite{ Joyez1997}
\begin{equation} \label{eq:energy}
E_{C}=g_{cl}\left(\frac{E_{11}}{g_L}+\frac{E_{21}}{g_m}+\frac{E_{22}}{g_R}\right)\, .
\end{equation}
The path average in Eq.\,(\ref{average}) and the partition function in Eq.\,(\ref{Zeff}) were expressed as the sum over paths with different boundary conditions. Instead of evaluating each path integral separately up to a certain cut-off and adding them up, it is more convenient to make the transformations
\begin{equation}\label{winding}
\varphi_{I}(\tau)=\xi_{I}(\tau)+\nu_{k_{I}}\tau,
\end{equation}
where $\nu_{k_{I}}= 2\pi k_{I}/(\beta E_{C})$ with the periodic path obeying $\xi_{I}(0)=\xi_{I}(\beta E_{C})$. When inserting the representation in Eq.\,(\ref{winding}) into the path average in Eq.\,(\ref{average}), we can rewrite the average as a path integral over periodic paths as
\begin{equation}\label{average2}
\left\langle\,  F[\vec{\xi}]\,\right\rangle
= \frac{1}{Z_{\rm eff}}  \oint D[\vec{\xi}] \sum_{k_{L},k_{R}=-\infty}^{\infty} {\rm e}^{-S_{{\rm eff},\vec{k}}[\vec{\xi}]}\, 
F_{\vec{k}}[\vec{\xi}]
\end{equation}
with
\begin{equation}\label{Zeffnew}
Z_{\rm eff}= \oint D[\vec{\xi}] \sum_{k_{L},k_{R}=-\infty}^{\infty} {\rm e}^{-S^{}_{{\rm eff},\vec{k}}[\vec{\xi}]}\, ,
\end{equation}
where $\vec{\xi}$ and $\vec{k}$ denote the periodic phase vector and the winding number vector, respectively. In units of $E_C$, the effective action in Eq.\,(\ref{effactionphi}) can be rewritten as
\begin{equation}\label{Seffk}
S_{{\rm eff},\vec{k}}[\vec{\xi}]=S_{C,\vec{k}}[\vec{\xi}]+ S^{}_{{\rm tun},\vec{k}}[\vec{\xi}]\, ,
\end{equation}
where the Coulomb action
\begin{equation}\label{SC0k}
S_{C,\vec{k}} [\vec{\xi}]= \frac{4\pi^2}{\beta E_{C}} \vec{k}^{T}\mathbb{E}\vec{k} + \int_{0}^{\beta E_{C}} d\tau\,
\dot{\vec{\xi}}^{T}\mathbb{E}\,\dot{\vec{\xi}}
\end{equation}
with the matrix
\begin{equation}\label{matrixC}
   \mathbb{E}=\frac{E_{C}}{2 e^{2}}\left(\begin{array}{cc} C_{\Sigma 1} & -C_{21}\\ -C_{21}& C_{\Sigma 2}
\end{array} \right).
\end{equation}
The tunnelling action reads 
\begin{eqnarray} \label{STun0k}
S_{{\rm tun},\vec{k}}[\vec{\xi}] &=&- \int_0^{\beta E_{C}}d\tau\int_0^{\beta E_{C}}d\tau^{\prime}\,\alpha(\tau-\tau^{\prime})
\Bigg(g_L\cos\left[\xi_1(\tau)-\xi_1(\tau^{\prime})+\nu_{k_{1}}(\tau-\tau^\prime)\right]\nonumber\\
&&+g_M\cos\left[\xi_1(\tau)-\xi_2(\tau) -\xi_1(\tau^{\prime})+\xi_2(\tau^{\prime})
+(\nu_{k_{1}}-\nu_{k_{2}})(\tau-\tau^\prime) \right]\nonumber\\
&&+g_2\cos\left[\xi_2(\tau)-\xi_2(\tau^{\prime})+\nu_{k_{2}}(\tau-\tau^\prime)\right]\Bigg)\,,
\end{eqnarray}
with
\begin{eqnarray}
\label{alphadimsionless}
\alpha(\tau-\tau^{\prime}) = \left[4 (\beta E_C)^2\sin^2\left(\frac{\pi(\tau-\tau^{\prime})}{\beta E_C}\right)\right]^{-1}\, .
\end{eqnarray}
Therefore, we performed the Monte Carlo sampling for the path variables ${\xi_1}$ and ${\xi_2}$, and the winding numbers $k_1$ and $k_2$. 
\section{Linear Response Conductance of SEP}
\label{sec:LRC}
According to the Kubo formula and following the idea used to calculate the DC conductance of the SET  \cite{Wallisser2002,Christoph2004}, the SEP's DC conductance can be expressed in terms of the Fourier transform of the real-time current correlation function $\bar C(t)$. However, this section briefly presents how to express the DC conductance in terms of the imaginary-time autocorrelation function $A(\tau)$. Using the definition in Eq.\,(\ref{FFphi}) and the convolution theorem of the Fourier transform, we obtain
\begin{eqnarray}
\label{eq:Geval0}
G &=& \lim_{\omega\to 0} \frac{\beta E_C}{2} \mathbf{\widetilde{C}}(\omega) \nonumber\\
&=& \beta E_C G_{0} \int_{-\infty}^\infty\!\! d\omega' \;\widetilde{\alpha}(-\omega') \; \tilde{A}(\omega')\, ,
\end{eqnarray}
where $\mathbf{\widetilde{C}}(\omega) =\int_{-\infty}^\infty\!\! d t  e^{-\,i \omega t}\bar C(t)$. With the substitution $\tau=it$, a Fourier transform of the tunnelling kernel $\alpha(\tau)$ is $\widetilde{\alpha}(\omega) = \omega/(2\pi(1-e^{- \beta E_C \omega}))$. The spectral function $\widetilde{A}(\omega)$ can be determined  from the autocorrelation function as  
\begin{eqnarray}
\label{eq:cosineIP0}
A(\tau) &=&\frac{1}{2\pi}\int_{-\infty}^\infty \!\! d\omega \; e^{-\tau \omega}\tilde{A}(\omega)\nonumber\\
&=& \frac{1}{2 \pi} \int_0^\infty d\omega \;\left[ e^{-\tau \omega} + e^{-(\beta E_C - \tau) \omega} \right]\; \tilde{A}(\omega),
\end{eqnarray}
where the detailed balance relation for the spectral function, i.e., $\tilde{A}(-\omega) = \exp(-\beta E_C \omega) \tilde{A}(\omega)$ has been applied. For numerical evaluation, it is more suitable to introduce the symmetric spectral function
\begin{equation}
\tilde{A}^s(\omega) \equiv \frac{1 - e^{-\beta E_{C} \omega}}{\omega} \tilde{A}(\omega)\, ,
\end{equation}
for which $\tilde{A}^s(-\omega) = \tilde{A}^s(\omega)$. In summary, Eq.\,(\ref{eq:Geval0}) and Eq.\,(\ref{eq:cosineIP0}) can be rewritten in terms of $\tilde{A}^s(\omega)$ as
\begin{eqnarray}
\label{eq:sepG}G &=& \frac{\beta E_C G_{0}}{2 \pi} \int_0^\infty
\!\! d\omega \; \frac{\omega^2}{\cosh(\beta E_C \omega) - 1} \;
\tilde{A}^s(\omega), \\[0.4cm]
\label{eq:setip} A(\tau) &=&\int_0^\infty \!\! d\omega \; \frac{\omega \cosh\left( \left[
\frac{\beta E_C}{2} - \tau \right] \omega \right)}{2 \pi
\sinh\left( \frac{\beta E_C}{2} \omega \right)} \; \tilde{A}^s(\omega).
\end{eqnarray}
To determine the DC conductance in Eq.\,(\ref{eq:sepG}), we first performed the PIMC method to calculate the autocorrelation function $A(\tau)$. Next, the inverse problem was numerically solved for the spectral function $\tilde{A}^s(\omega)$ based on the single–value decomposition (SVD) for the integral operator \cite{Wallisser2002,hansen87,louis89,press92}. 

\section{Results and Discussion}
\label{sec:results}

Before presenting the results for the autocorrelation function and the SEP's conductance, we introduce the relevant Monte Carlo parameters corresponding to the SEP's experiment. Although the experiment operated in the temperature range between $26\, \text{mK}$ and $ 20 \,\text{K}$, we only calculated the DC conductance in the temperature range between $100\, \text{mK}$ and $ 20 \,\text{K}$ due to the influence of the fermionic sign problem \cite{Ceperley1995, Troyer2005}. Since the experiment and our theoretical calculation had chosen the difference in charging energy, we then presented the results in terms of the temperature using the relationship $T = E_C/k_{B} $, where $E_C = 0.184\, \text{meV} $ was obtained by Eq.\,(\ref{eq:energy}). The correlation function for each parameter was calculated using the difference in the Monte Carlo measurements to obtain a statistical error of less than $ 1\%$. Furthermore, the PIMC simulations were performed at each temperature to obtain the number of correlation functions, which were adjusted to ensure that the statistical error of the DC conductance was below $1\%$ over the whole temperature range, except for the lowest temperature, where the statistical error of the maximal conductance is about $5\%$.

\begin{figure}
\centering
\includegraphics[width= 1.0\textwidth]{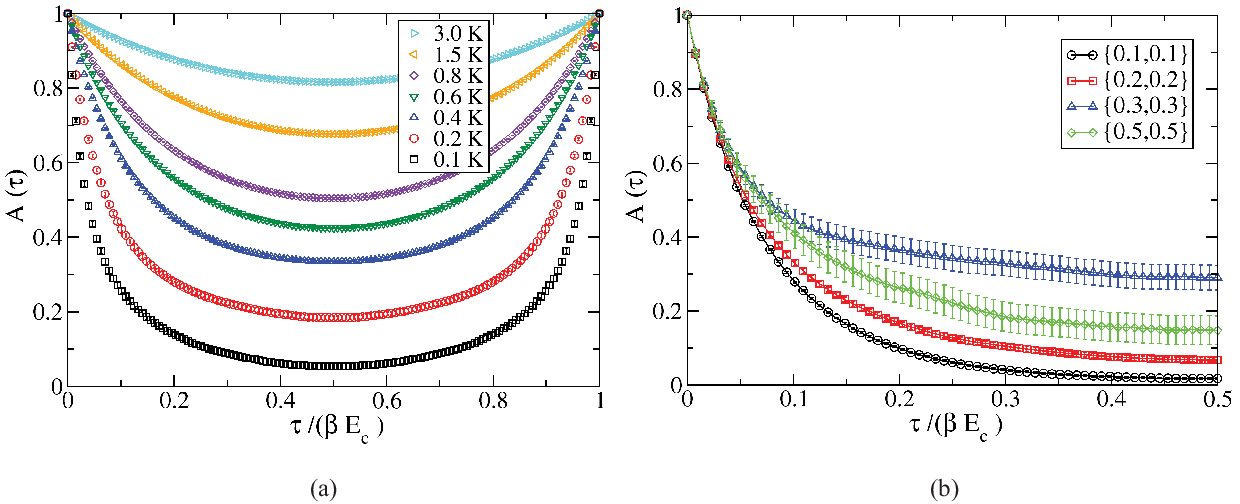}
\vspace{0.2cm}
\caption{ (a) PIMC results for the imaginary--time correlation functions $A(\tau)$ of the SEP for different temperatures. Error bars denote one standard deviation and are smaller than the symbol size. (b) The imaginary--time correlation functions with different values of the gate voltages at temperature $ T = 0.2\text{K}$.}
\label{fig:cf}  
\end{figure}

Fig.\,\ref{fig:cf} (a) illustrates the autocorrelation functions when both gate voltages equal $0.0$ for varying temperatures. The findings indicate that the temperature significantly impacts the amplitude of $A(\tau)$. Furthermore, the autocorrelation function's dependence on the gate voltages is displayed in Fig.\,\ref{fig:cf} (b). The autocorrelation function's statistical errors strongly depend on the gate voltages. They are commonly at their least apparent at $\vec{n}_{g}=0$, but as the gate voltages increase, they become more evident until they reach their maximum at $ \{n_{01},n_{02}\} =\{0.5,0.5\}$, shown with the green dots in Fig.\,\ref{fig:cf} (b). The relative statistical error between the minimum and maximum case is about $20\%$  at temperature $ T = 0.2\text{K}$. The signal-to-noise ratio increases as the gate voltages increase, reflecting the fermionic sign problem. 

\begin{figure}
\centering
\includegraphics[width=0.75\textwidth]{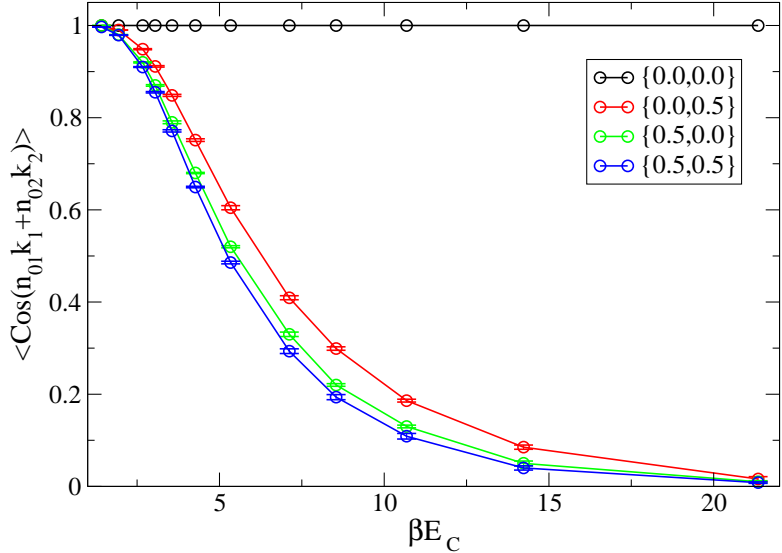}
\vspace{0.2cm}
\caption{ The average sign term as a function of $\beta E_C$ illustrates the impact of the fermionic sign problem in the SEP. The sign term remains unity (black line) for the absence of the gate voltages, indicating no sign problem. However, cancellations in the sign term lead to significant suppression at low temperatures due to the gate voltage presenting $\{n_{01},n_{02}\}$.}
\label{fig:sign_problem}
\end{figure}

Furthermore, we focused on the influence of the fermionic sign problem increasing at low temperatures by analyzing through the oscillatory factor in Eq.\,(\ref{correlation2}). The sign term defined as the average $\langle \cos(n_{01}k_{1}+ n_{02}k_{2}) \rangle $ is calculated and shown in Fig.\,\ref{fig:sign_problem}, where the sine term vanishes because the average of an odd function is zero.  As a result, the average of the sign term remains in unity due to the absence of gate voltage, as the black line indicates in this figure. However, for the gate voltage present, the cancellation in the sign term causes its average to diminish significantly. Consequently, the average in Eq.\,(\ref{average}), which is determined as a ratio of the sign term, is dominated by noise and becomes inaccurate in the quantum Monte Carlo simulation. In addition, due to the asymmetry of the SEP, the average sign term differs for the cases $\{ 0.0, 0.5\} $ and $\{ 0.5, 0.0\}$ plotted as the red and green lines. 

\begin{figure}
\centering
\includegraphics[width=0.75\textwidth]{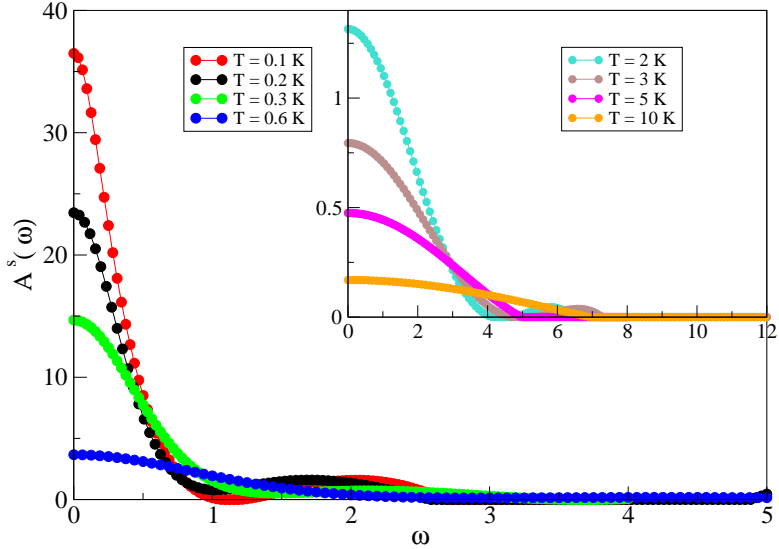}
\vspace{0.2cm}
\caption{ Symmetric spectral function $\tilde{A}^s(\omega)$  for the gate voltage $\vec{n}_g=0$. These symmetric spectral functions were calculated with the upper cut-off frequency $\omega_{\max}=4$ and $\omega_{\max}=8$. However, in both cases, the amplitudes of $\tilde{A}^s(\omega)$ become zero for $\omega > 4$. Therefore, it is reasonable to introduce the cut-off frequency.}
\label{fig:sepSPEC}
\end{figure}

From the solution of the inverse problem in Eq.\,(\ref{eq:setip}), we obtained the symmetric spectral function and then determined the DC conductance of the SEP according to Eq.\,(\ref{eq:sepG}). However, an upper cut-off frequency must be determined to estimate the integration for more convenience numerically. For example, the results for the symmetric spectral function $\tilde{A}^s(\omega)$ are shown as a frequency function for the absence of the gate voltages in Fig.\,\ref{fig:sepSPEC}. Obviously, the amplitude of $\tilde{A}^s(\omega)$ strongly decays with increasing frequency. Consequently, we introduced an upper cut-off frequency $\omega_{\max}$ to numerically calculate the integral in Eq.\,(\ref{eq:sepG}). As shown in Fig.\,\ref{fig:sepSPEC}, at low-temperature $(T \ll E_C /k_B)$, we used the frequency cut-off $\omega_{\max}=4.0$ for the numerical DC conductance evaluation. Likewise, for high-temperature $(T \geq E_C /k_B)$, the magnitude of  $\tilde{A}^s(\omega)$ slowly decays with increasing frequency, such that the frequency cut-off was chosen as $\omega_{\max}= 10$. 

\begin{figure}
\centering
\includegraphics[width=0.90\textwidth]{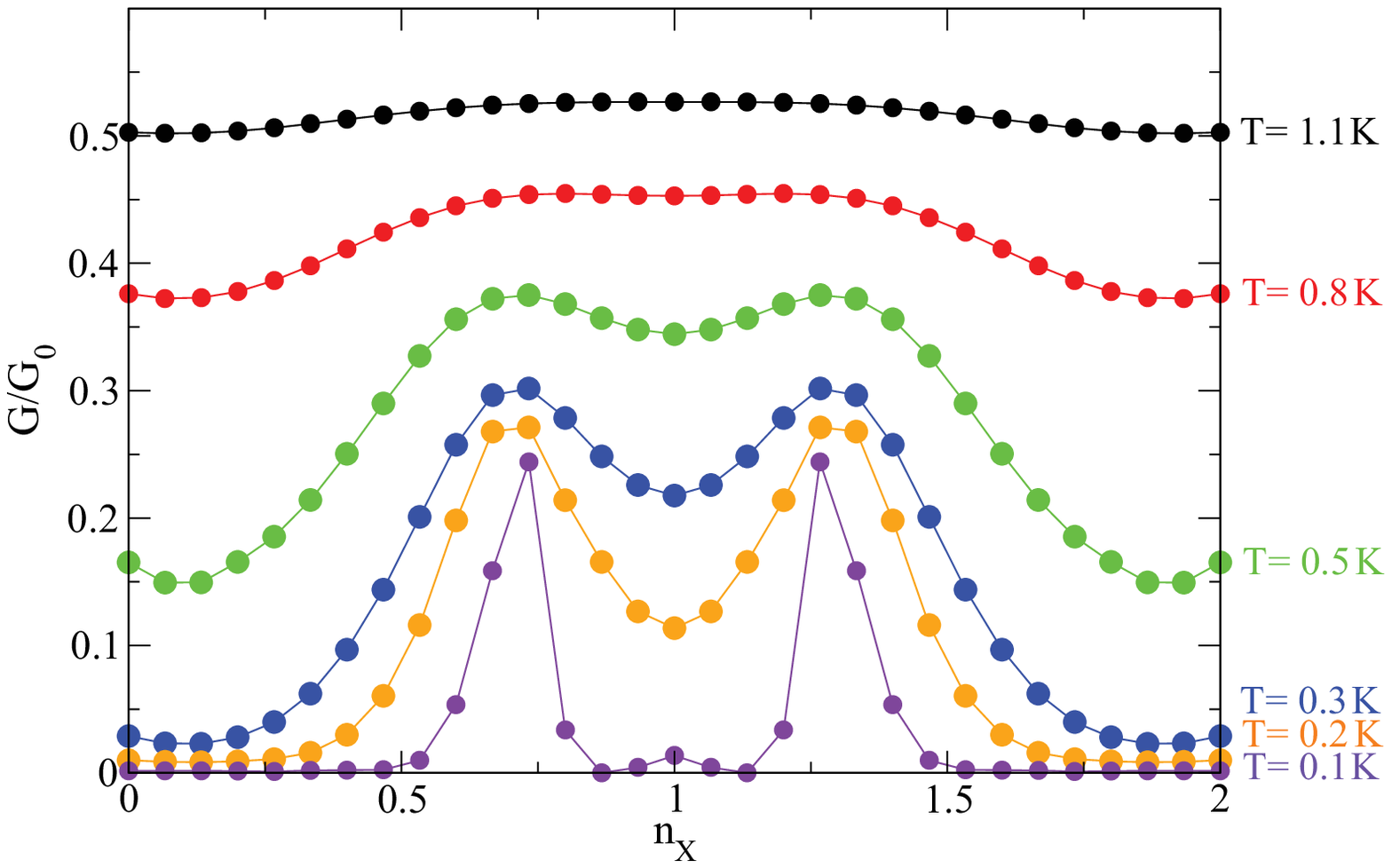}
\vspace{0.2cm}
\caption{ Coulomb oscillations of the DC conductance of the SEP for different temperatures from $0.1 K$ (lowest curve) to $ 1.1\,K $ (top curve).}
\label{fig:Coulomb_osc.jpg}
\end{figure}

The DC conductance depends on the temperature and two gate voltages since the autocorrelation function depends on both parameters. To calculate the Coulomb oscillations of the conductance for various temperatures, we focused on the gate voltage condition plotted by the red dot line in Fig.\ref{fig:SEP}\,(b), which is the boundary of the honeycombs of the states $(0,1)$ and $(1,0)$, and passes two classical triple points. By varying $n_{01}\in \{0,1\}$, the corresponding dimensionless gate voltage $n_{02}$ was calculated by Eq.\,(\ref{condition_6}) to perform the PIMC simulation. Fig.\ref{fig:Coulomb_osc.jpg} shows Coulomb oscillations of the Conductance for various $n_{x}=n_{01}+n_{02}$. As a result, for low-temperature $T\ll 2.13\,\text{K} $, two conductance peaks are apparent in the Coulomb oscillations where the conductance reaches a maximum at the triple points. These results reflect that electrons transferring through the SEP correspond with the electron states in the sequence $(0,0) \rightarrow (1,0) \rightarrow (0,1)$ for the first peak and $(1,0) \rightarrow (0,1) \rightarrow (1,1)$ for the second peak. However, the Coulomb blockade effect vanishes at high-temperature $T\gg 2.13\,\text{K} $, and the Coulomb oscillation is smeared out because electrons can transfer continuously through the system.

\begin{figure}
\centering
\includegraphics[width=0.75\textwidth]{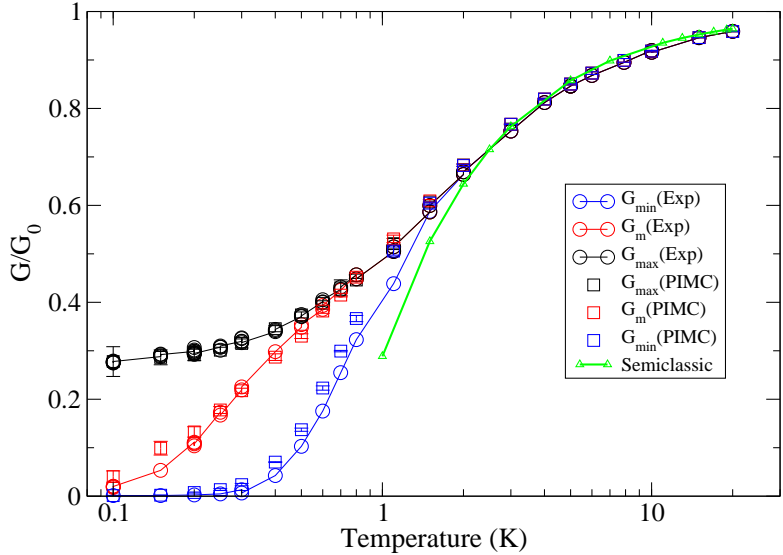}
\vspace{0.2cm}
\caption{Minimal middle and maximal conductance of the SEP as a function of temperature compared with the experimental data.}
\label{fig:fig_con}
\end{figure}

For low temperatures $(T \ll E_C /k_B)$, the DC conductance depends on the gate voltages. It is better to compare the experimental and theoretical results with three types of conductance that occur for different values of $n_{01}$ and $n_{02}$. When setting $n_{01}$ and $n_{02}$ to zero, the minimal conductance $ G_{\text{min}} $ was calculated and compared with the experimental data, as indicated in blue in Fig.\,\ref{fig:fig_con}. Since the states beside $(0,0)$ are occupied only at high temperatures, $ G_{\text{min}} $ increases to merge with other conductance. However, $ G_{\text{min}} $ is exponentially small for low temperatures. The small minimal conductance is due to the thermal fluctuation of the background charge. The middle conductance $G_{\text{m}}$ was calculated by fixing $n_{01}=0.5$ and $n_{02}= 0.5$, as shown with the red square marks in Fig.\,\ref{fig:fig_con}. As a result, the linear decay in the semi-logarithmic plot demonstrates that the conductance is exponentially suppressed because the charges fluctuate between two states, $(1,0)$ and $(0,1)$, at a low temperature. However, the states $(0,0)$ and $(1,1)$ are thermally occupied, leading to an exponential increase in $G_{\text{m}}$, which finally approaches $ G_{0}$ at high-temperature.

Since only the classical triple points are indicated as blue circles in Fig.\ref{fig:SEP}\, (b), they are identified in the stability diagram calculated with the absolute zero temperature condition, the position of the maximal conductance is unknown and depends on the temperature \cite{Limbach2005}. In the experiment to determine the maximal conductance of the SEP at low temperatures, a $6\times 6$ conductance grid was first measured in the $(n_{01},n_{02})$ plane. Subsequently, the maximal conductance and its position regarding the gate voltages were identified using a Gaussian fit. Similarly, we performed the PIMC simulations and SVD calculations for each temperature for the theoretical calculation, with various gate voltages on the $6\times 6$ grid for $n_{01}\in \{0.0, 0.5\}$ and $n_{02} \in \{0.0, 0.5\}$. We observed that the conductance reaches its maximum in the ranges of $0.3 < n_{01} < 0.5$ and $0.2 < n_{02} < 0.4$. The Gaussian fit was then used to determine the maximal conductance at each temperature, expressed as 
\begin{equation}
\label{eq:Gfit}
G(n_{01}, n_{02})=G_{\text{max}}e^{-\mathcal{A}_{1} (n_{01}- n_{01}^{\text{max}})^2-\mathcal{A}_{2} (n_{01}- n_{01}^{\text{max}})(n_{02}- n_{02}^{\text{max}})-\mathcal{A}_{3} (n_{02}- n_{02}^{\text{max}})^2}\, ,
\end{equation}
where $G_{\text{max}}$, $n_{01}^{\text{max}}$, $n_{02}^{\text{max}}$, $\mathcal{A}_{1}$, $\mathcal{A}_{2}$ and $\mathcal{A}_{3}$ are unknown. The maximal conductance $G_{\text{max}}$ is obtained at position $(n_{01}^{\text{max}},n_{02}^{\text{max}})$, called the quantum triple point. For low temperatures, this position is different from the classical triple point. This result corresponds with SEP's experiment \cite{Limbach2005}. The experimental and theoretical maximal conductance comparison was plotted in black in Fig.\,\ref{fig:fig_con}. The results show that highly accurate PIMC results can describe the experimental maximal conductance. The three conductance types of the SEP are different at temperatures below $3\text{K}$. However, they are essentially the same at temperatures above $3\text{K}$. In essence, this shows the Coulomb blockade effect in the two-island system. 

For high-temperature $T \gg E_C /k_B$, the DC conductance of the SEP does not depend on the gate voltages. It can be calculated by the semiclassical approximation, i.e., $G_{\text{cl}} \approx G_{0}(1-\beta E_{C}/3)$ \cite{Limbach2005,Joyez1997}. The PIMC conductance converges to $G_{\text{cl}}$, as shown by the green line in Fig.\,\ref{fig:fig_con}. The semiclassical approximation exhibits a more significant deviation when the temperature becomes much lower than $3\text{K}$. These results confirm that the semiclassical approach is valid only at high temperatures. In contrast, the PIMC results agree with the experimental data throughout the range of temperatures. However, for temperatures lower than those shown in Fig.\,\ref{fig:fig_con}, converging Monte Carlo results in a reasonable time were impossible because of the influence of the fermionic sign problem.

\section{Conclusions}
\label{sec:conclusion}
In this Letter, we calculated the SEP's conductance using the PIMC method and the SVD to solve the inverse problem. Furthermore, we compared the theoretical results with the experimental results. The minimal, middle, and maximal conductance results match nicely even in the low-temperature regime where strong quantum fluctuation occurs. Alternatively, the semiclassical approach can only be used to describe the experimental data in the high-temperature regime accurately. Without such a temperature limitation, the PIMC approach has a significant advantage, especially when describing the quantum fluctuation phenomena of the SEP. Therefore, this study repeats the PIMC method's success in accurately describing the quantum fluctuation phenomena of single-electron devices. Finally, our findings leave ample space for further work. The term involves a multiplication of two current functionals, giving no contribution to the current correlation function of the SEP by the numerical proof. However, it would be possible to give some analytical proof. Moreover, since the statistical error of the Monte Carlo method increases exponentially with inverse temperature, limitations of the Monte Carlo calculation exist at very low temperatures. However, P. Werner \textit{et al.}\,\cite{Lukyanov2006} reported that the cluster Monte Carlo algorithm can be applied to study a single electron box even at very low temperatures. Further work is needed to investigate whether this algorithm also has advantages in calculating the correlation function of SEPs.

\section*{Acknowledgments}
This research project was financially supported by Mahasarakham University for the main funding and the NanoMaterials Physics Research Unit (NMPRU). We especially thank C.Theis and L. Mühlbacher for valuable discussions.
\appendix
\section{ Stability Diagram}
\label{sec:SD}
Following the ideal in Ref.\cite{ Fujisawa2002}, the stability diagram of the SEP can be constructed using the charging energy expressed in Eq.(\ref{Ec0}). In the absence of the applied voltage $V_{DS}=0$, the electrochemical potential of the SEP must be zero. The six border lines of a honeycomb can be calculated from the fact that an electron can tunnel in six distinct ways by applying two gate voltages. From the electrochemical potential definition, the six conditions read as
\begin{equation}\label{Ecn1}
E_C(n_1-1,n_2)=E_C(n_1,n_2),
\end{equation}
\begin{equation}\label{Ecn2}
    E_C(n_1+1,n_2)=E_C(n_1,n_2),
\end{equation}
\begin{equation}\label{Ecn3}
   E_C(n_1,n_2-1)=E_C(n_1,n_2),
\end{equation}
\begin{equation}\label{Ecn4}
    E_C(n_1,n_2+1)=E_C(n_1,n_2),
\end{equation}
\begin{equation}\label{Ecn5}
    E_C(n_1+1,n_2-1)=E_C(n_1,n_2),
\end{equation}
and,
\begin{equation}\label{Ecn6}
    E_C(n_1-1,n_2+1)=E_C(n_1,n_2).
\end{equation}
Inserting the charging energy in Eq.(\ref{Ec0}) into Eqs.(\ref{Ecn1}--\ref{Ecn6}), one obtains six linear equations as a function of two gate variables and excess charge numbers. By fixing the values of $n_{1}$  and $n_{2}$, the intersection of the six linear equations produces a hexagonal area of $(n_{1},n_{2})$. However, we focused only on the case where an electron transfers from the left to right islands, corresponding with Eq.(\ref{Ecn6}) expressed as 
\begin{eqnarray}
\label{condition_6}
  n_{02} =  \bigg({\frac{C_{21} - C_{\Sigma 2}}{C_{21} - C_{\Sigma 1}}}\bigg){n_{01}} +\bigg(\frac{3 C_{\Sigma 1} +C_{\Sigma 2}-2C_{21}}{2({C_{21} - C_{\Sigma 1}})}\bigg)  \, ,
\end{eqnarray}
where we fixed $n_{1}=0$ and $n_{2}=1$. This linear equation was plotted by the red dot line in Fig.\ref{fig:SEP} (b). 

\section{Generating Functional of SEP}
\label{sec:GF_SEP}
Starting with the path integral representation of the generating functional $Z_{\rm gen}[\rchi]$ of the SEP system, including the auxiliary fields $\rchi$ \cite{PhysRevB.58.R10155}, 
\begin{equation}\label{Zgen}
 Z_{\rm gen}[\rchi] = {\rm tr}\, T_{\tau}\{{\rm e}^{-\beta (\hat{H}-\sum_{K} \hat{I}_{K} \rchi_{K}(\tau))}\},
\end{equation}
where $  T_{\tau}$ is the time ordering operator for imaginary time $\tau$, and $\hat{H}$ is the Hamiltonian of the system expressed in Eq.\,(\ref{Hamiltonian}). The tunnelling currents through the left $(L)$, middle $(M)$, and right $(R)$ junctions correspond to the current operators
 \begin{eqnarray}\label{ISEP}
  \hat{I}_L(\tau)&=&ie\sum_{kq\sigma} \left(t_{1Skq \sigma}\, {\rm
  e}^{-i\varphi^{}_{1}}d_{1k\sigma}^{\dag}c_{Sq\sigma}^{}-\hbox{H.c.}\right)\nonumber\\
   \hat{I}_M(\tau)&=&ie\sum_{kq\sigma}  \left(t_{21kq\sigma}\, {\rm
   e}^{i(\varphi_{1}^{}-\varphi_{2}^{})}d_{2k\sigma}^{\dag}d_{1q\sigma}^{}-\hbox{H.c.}\right)\nonumber \\
 \hat{I}_R(\tau)&=&ie\sum_{k,q,\sigma}\left(t_{D2kq \sigma}\, {\rm
 e}^{i\varphi_{2}}c_{Dk\sigma}^{\dag}d_{2q\sigma}^{}-\hbox{H.c.}\right),
\end{eqnarray}
respectively, where electrons tunnel from source to drain. According to the representation in Eq.\,(\ref{Zgen}), $Z_{\rm gen}[\chi]$ may be written as
\begin{eqnarray}\label{Zchi3}
Z_{\rm gen}[\chi]&=& \mathcal{N} \sum_{k_{1},k_{1}=-\infty}^{\infty}
\int\limits_{\varphi_{1}(0)}^{\varphi_{1}(0)+2\pi k_{1}}\!\!\!\!
D[\varphi_{1}]\!\!\int\limits_{\varphi_{2}(0)}^{\varphi_{2}(0)+2\pi k_{2}}\!\!\!\! D[\varphi_{2}]\nonumber \\ &&\quad\times{\rm
e}^{-2\pi i\,(\vec{n}_{g}\cdot\vec{k})}\, {\rm e}^{-S_C[\vec{\varphi}]- S_{\rm tun}[\vec{\varphi},\chi]} \, ,
\end{eqnarray}
where $\mathcal{N}$ is just the normalized factor corresponding with the partition function of the (isolated) leads and islands. For the paths $\vec{\varphi}(\tau)=(\varphi_1(\tau),\varphi_2(\tau))^{T}$  with given winding numbers $\vec{k}=(k_1,k_2)^{T}$, the Coulomb action in Eq.\,(\ref{Zchi3}) reads
\begin{equation}
\label{SC2}
S_{C} [\vec{\varphi}]= \frac{1}{4}\int_{0}^{\beta} d\tau\,
\vec{\dot{\varphi}}^{T}\widetilde{\mathbb{E}}\,\vec{\dot{\varphi}} +2\pi i\,(\vec{n}_{g}\cdot\vec{k}),
\end{equation}
where the matrix is $\widetilde{\mathbb{E}}=\mathbb{E}_{C}^{-1}$. The tunnelling action in  Eq.\,(\ref{Zchi3}) takes the form 
\begin{equation}
\label{SsubTun4}
S_{\rm tun}[\vec{\varphi},\chi] =S^{}_{\rm tun}[\vec{\varphi}] +S^{(1)}_{\rm tun}[\vec{\varphi},\chi]
+ S^{(2)}_{\rm tun}[\vec{\varphi},\chi]\, ,
\end{equation}
where
\begin{eqnarray}
\label{STun0}
S^{}_{\rm tun}[\vec{\varphi}] &=& -
\int_0^{\beta}d\tau\int_0^{\beta}d\tau^{\prime}\,\alpha(\tau-\tau^{\prime})\nonumber\\
&&\quad\times\bigg(g^{}_{L}\cos\left[\varphi_1(\tau)-\varphi_1(\tau^{\prime})\right]
+g_{R}\cos\left[\varphi_2(\tau)-\varphi_2(\tau^{\prime})\right]\nonumber\\
& &\quad\quad+g_{M}\cos\left[\varphi_1(\tau)-\varphi_2(\tau)-\varphi_1(\tau^{\prime})+\varphi_2(\tau^{\prime})\right]\bigg)\,
\end{eqnarray}
is the tunnelling action in the absence of the source term. The tunnelling kernel $\alpha(\tau)$ describing the electron-hole pair propagation for electrons and holes created in different electrodes can be represented in the form  
\begin{eqnarray}
\alpha(\tau-\tau^{\prime}) =\frac{1}{4\beta^2 \sin^2\left(\frac{\pi(\tau-\tau^{\prime})}{\beta }\right)} \,  .
\end{eqnarray}
The linear and quadratic source terms take the from
\begin{eqnarray} 
\label{STun1}
S^{(1)}_{\rm tun}[\vec{\varphi},\chi]&=&2e\int_0^{\beta}d\tau\int_0^{\beta}d\tau^{\prime}\,\alpha(\tau-\tau^{\prime})\nonumber\\
&&\times\Big(\,g_{L}\chi_L(\tau)\sin\left[\varphi_1(\tau)-\varphi_1(\tau^{\prime})\right]
-g_{R}\chi_R(\tau)\sin\left[\varphi_2(\tau)-\varphi_2(\tau^{\prime})\right]\nonumber\\ &&\ -
g_M\chi_M(\tau)\sin\left[\varphi_1(\tau)-\varphi_2(\tau)-\varphi_1(\tau^{\prime})+\varphi_2(\tau^{\prime})\right]\Big)\,
\end{eqnarray}
and
\begin{eqnarray}
\label{STun2}
S^{(2)}_{\rm tun}[\vec{\varphi},\chi] &=&-e^2\int_0^{\beta}d\tau\int_0^{\beta}d\tau^{\prime}\,\alpha(\tau-\tau^{\prime})
\Big(\,g_{L}\chi_L(\tau)\chi_L(\tau^{\prime})\cos\left[\varphi_1(\tau)-\varphi_1(\tau^{\prime})\right]\nonumber \\ &&\ +
g^{}_{M}\chi_M(\tau)\chi_M(\tau^{\prime})\cos\left[\varphi_1(\tau)-\varphi_2(\tau)-\varphi_1(\tau^{\prime})
+\varphi_2(\tau^{\prime})\right]\Big)\nonumber\, \\
&& +g_{R}\chi_R(\tau)\chi_R(\tau^{\prime})\cos\left[\varphi_2(\tau)-\varphi_2(\tau^{\prime})\right] .
\end{eqnarray}

\bibliographystyle{model1a-num-names}
\bibliography{REF}

\begin{thebibliography}{46}
\expandafter\ifx\csname natexlab\endcsname\relax\def\natexlab#1{#1}\fi
\providecommand{\bibinfo}[2]{#2}
\ifx\xfnm\relax \def\xfnm[#1]{\unskip,\space#1}\fi
\bibitem[{Ploog(1993)}]{Grabert}
\bibinfo{author}{K.~H. Ploog}, \bibinfo{journal}{Advanced Materials} \bibinfo{volume}{5} (\bibinfo{year}{1993}) \bibinfo{pages}{227--227}.
\bibitem[{Howe et~al.(2021)Howe, Blumenthal, Beere, Mitchell, Ritchie, and Pepper}]{Howe2021}
\bibinfo{author}{H.~Howe}, \bibinfo{author}{M.~Blumenthal}, \bibinfo{author}{H.~E. Beere}, \bibinfo{author}{T.~Mitchell}, \bibinfo{author}{D.~A. Ritchie}, \bibinfo{author}{M.~Pepper}, \bibinfo{journal}{Applied Physics Letters} \bibinfo{volume}{119} (\bibinfo{year}{2021}) \bibinfo{pages}{153102}.
\bibitem[{Milton et~al.(2010)Milton, Williams, and Forbes}]{Milton_2010}
\bibinfo{author}{M.~J.~T. Milton}, \bibinfo{author}{J.~M. Williams}, \bibinfo{author}{A.~B. Forbes}, \bibinfo{journal}{Metrologia} \bibinfo{volume}{47} (\bibinfo{year}{2010}) \bibinfo{pages}{279}.
\bibitem[{Pekola et~al.(2013)Pekola, Saira, Maisi, Kemppinen, M\"ott\"onen, Pashkin, and Averin}]{RevModPhys.85.1421}
\bibinfo{author}{J.~P. Pekola}, \bibinfo{author}{O.-P. Saira}, \bibinfo{author}{V.~F. Maisi}, \bibinfo{author}{A.~Kemppinen}, \bibinfo{author}{M.~M\"ott\"onen}, \bibinfo{author}{Y.~A. Pashkin}, \bibinfo{author}{D.~V. Averin}, \bibinfo{journal}{Rev. Mod. Phys.} \bibinfo{volume}{85} (\bibinfo{year}{2013}) \bibinfo{pages}{1421--1472}.
\bibitem[{Giblin et~al.(2012)Giblin, Kataoka, Fletcher, See, Janssen, Griffiths, Jones, Farrer, and Ritchie}]{Giblin_2012}
\bibinfo{author}{S.~Giblin}, \bibinfo{author}{M.~Kataoka}, \bibinfo{author}{J.~Fletcher}, \bibinfo{author}{P.~See}, \bibinfo{author}{T.~Janssen}, \bibinfo{author}{J.~Griffiths}, \bibinfo{author}{G.~Jones}, \bibinfo{author}{I.~Farrer}, \bibinfo{author}{D.~Ritchie}, \bibinfo{journal}{Nature Communications} \bibinfo{volume}{3} (\bibinfo{year}{2012}) \bibinfo{pages}{930}.
\bibitem[{Harrington et~al.(2022)Harrington, Mueller, and Murch}]{Harrington2022}
\bibinfo{author}{P.~M. Harrington}, \bibinfo{author}{E.~J. Mueller}, \bibinfo{author}{K.~W. Murch}, \bibinfo{journal}{Nature Reviews Physics} \bibinfo{volume}{4} (\bibinfo{year}{2022}) \bibinfo{pages}{660–671}.
\bibitem[{DeMille et~al.(2024)DeMille, Hutzler, Rey, and Zelevinsky}]{DeMille2024}
\bibinfo{author}{D.~DeMille}, \bibinfo{author}{N.~R. Hutzler}, \bibinfo{author}{A.~M. Rey}, \bibinfo{author}{T.~Zelevinsky}, \bibinfo{journal}{Nature Physics} \bibinfo{volume}{20} (\bibinfo{year}{2024}) \bibinfo{pages}{741–749}.
\bibitem[{Pashkin et~al.(2003)Pashkin, Yamamoto, Astafiev, Nakamura, Averin, and Tsai}]{Pashkin2003}
\bibinfo{author}{Y.~A. Pashkin}, \bibinfo{author}{T.~Yamamoto}, \bibinfo{author}{O.~Astafiev}, \bibinfo{author}{Y.~Nakamura}, \bibinfo{author}{D.~V. Averin}, \bibinfo{author}{J.~S. Tsai}, \bibinfo{journal}{Nature} \bibinfo{volume}{421} (\bibinfo{year}{2003}) \bibinfo{pages}{823--826}.
\bibitem[{Pesin and Andreev(2004)}]{Andreev2004}
\bibinfo{author}{D.~A. Pesin}, \bibinfo{author}{A.~V. Andreev}, \bibinfo{journal}{Phys. Rev. Lett.} \bibinfo{volume}{93} (\bibinfo{year}{2004}) \bibinfo{pages}{196808}.
\bibitem[{Hines et~al.(2007)Hines, Jacobs, and Wang}]{Hines_2007}
\bibinfo{author}{C.~Hines}, \bibinfo{author}{K.~Jacobs}, \bibinfo{author}{J.~B. Wang}, \bibinfo{journal}{Journal of Physics A: Mathematical and Theoretical} \bibinfo{volume}{40} (\bibinfo{year}{2007}) \bibinfo{pages}{F609}.
\bibitem[{Yang et~al.(2011)Yang, Wang, and Das~Sarma}]{Shuo2011}
\bibinfo{author}{S.~Yang}, \bibinfo{author}{X.~Wang}, \bibinfo{author}{S.~Das~Sarma}, \bibinfo{journal}{Phys. Rev. B} \bibinfo{volume}{83} (\bibinfo{year}{2011}) \bibinfo{pages}{161301}.
\bibitem[{Utsugi et~al.(2023)Utsugi, Lee, Tsuchiya, Mine, Mizokuchi, Yoneda, Kodera, Saito, Hisamoto, and Mizuno}]{Utsugi_2023}
\bibinfo{author}{T.~Utsugi}, \bibinfo{author}{N.~Lee}, \bibinfo{author}{R.~Tsuchiya}, \bibinfo{author}{T.~Mine}, \bibinfo{author}{R.~Mizokuchi}, \bibinfo{author}{J.~Yoneda}, \bibinfo{author}{T.~Kodera}, \bibinfo{author}{S.~Saito}, \bibinfo{author}{D.~Hisamoto}, \bibinfo{author}{H.~Mizuno}, \bibinfo{journal}{Japanese Journal of Applied Physics} \bibinfo{volume}{62} (\bibinfo{year}{2023}) \bibinfo{pages}{SC1020}.
\bibitem[{Blumenthal et~al.(2023)Blumenthal, Mahony, Ahmad, Gouveia, Howe, Beere, Mitchel, Ritchie, and Pepper}]{Blumenthal1a2023}
\bibinfo{author}{M.~D. Blumenthal}, \bibinfo{author}{D.~Mahony}, \bibinfo{author}{S.~Ahmad}, \bibinfo{author}{D.~Gouveia}, \bibinfo{author}{H.~Howe}, \bibinfo{author}{H.~E. Beere}, \bibinfo{author}{T.~Mitchel}, \bibinfo{author}{D.~A. Ritchie}, \bibinfo{author}{M.~Pepper}, \bibinfo{journal}{EPJ Quantum Technology} \bibinfo{volume}{10} (\bibinfo{year}{2023}).
\bibitem[{Pothier et~al.(1991)Pothier, Lafarge, Orfila, Urbina, Esteve, and Devoret}]{Pothier1991}
\bibinfo{author}{H.~Pothier}, \bibinfo{author}{P.~Lafarge}, \bibinfo{author}{P.~Orfila}, \bibinfo{author}{C.~Urbina}, \bibinfo{author}{D.~Esteve}, \bibinfo{author}{M.~Devoret}, \bibinfo{journal}{Physica B: Condensed Matter} \bibinfo{volume}{169} (\bibinfo{year}{1991}) \bibinfo{pages}{573--574}.
\bibitem[{Pothier et~al.(1992)Pothier, Lafarge, Urbina, Esteve, and Devoret}]{Pothier1992}
\bibinfo{author}{H.~Pothier}, \bibinfo{author}{P.~Lafarge}, \bibinfo{author}{C.~Urbina}, \bibinfo{author}{D.~Esteve}, \bibinfo{author}{M.~H. Devoret}, \bibinfo{journal}{Europhysics Letters ({EPL})} \bibinfo{volume}{17} (\bibinfo{year}{1992}) \bibinfo{pages}{249--254}.
\bibitem[{Limbach et~al.(2005)Limbach, vom Stein, Wallisser, and Sch\"afer}]{Limbach2005}
\bibinfo{author}{B.~Limbach}, \bibinfo{author}{P.~vom Stein}, \bibinfo{author}{C.~Wallisser}, \bibinfo{author}{R.~Sch\"afer}, \bibinfo{journal}{Phys. Rev. B} \bibinfo{volume}{72} (\bibinfo{year}{2005}) \bibinfo{pages}{045319}.
\bibitem[{Kaestner et~al.(2008)Kaestner, Kashcheyevs, Amakawa, Blumenthal, Li, Janssen, Hein, Pierz, Weimann, Siegner, and Schumacher}]{Kaestner2008}
\bibinfo{author}{B.~Kaestner}, \bibinfo{author}{V.~Kashcheyevs}, \bibinfo{author}{S.~Amakawa}, \bibinfo{author}{M.~D. Blumenthal}, \bibinfo{author}{L.~Li}, \bibinfo{author}{T.~J. B.~M. Janssen}, \bibinfo{author}{G.~Hein}, \bibinfo{author}{K.~Pierz}, \bibinfo{author}{T.~Weimann}, \bibinfo{author}{U.~Siegner}, \bibinfo{author}{H.~W. Schumacher}, \bibinfo{journal}{Phys. Rev. B} \bibinfo{volume}{77} (\bibinfo{year}{2008}) \bibinfo{pages}{153301}.
\bibitem[{Prada and Platero(2012)}]{Prada2012}
\bibinfo{author}{M.~Prada}, \bibinfo{author}{G.~Platero}, \bibinfo{journal}{Phys. Rev. B} \bibinfo{volume}{86} (\bibinfo{year}{2012}) \bibinfo{pages}{165424}.
\bibitem[{Tanttu et~al.(2015)Tanttu, Rossi, Tan, Huhtinen, Chan, Möttönen, and Dzurak}]{Tanttu2015}
\bibinfo{author}{T.~Tanttu}, \bibinfo{author}{A.~Rossi}, \bibinfo{author}{K.~Y. Tan}, \bibinfo{author}{K.-E. Huhtinen}, \bibinfo{author}{K.~W. Chan}, \bibinfo{author}{M.~Möttönen}, \bibinfo{author}{A.~S. Dzurak}, \bibinfo{journal}{New Journal of Physics} \bibinfo{volume}{17} (\bibinfo{year}{2015}) \bibinfo{pages}{103030}.
\bibitem[{van~der Heijden et~al.(2017)van~der Heijden, Tettamanzi, and Rogge}]{Joost2017}
\bibinfo{author}{J.~van~der Heijden}, \bibinfo{author}{G.~Tettamanzi}, \bibinfo{author}{S.~Rogge}, \bibinfo{journal}{Scientific Reports} \bibinfo{volume}{7} (\bibinfo{year}{2017}) \bibinfo{pages}{44371}.
\bibitem[{Schoinas et~al.(2024)Schoinas, Rath, Norimoto, Xie, See, Griffiths, Chen, Ritchie, Kataoka, Rossi, and Rungger}]{Schoinas2024}
\bibinfo{author}{N.~Schoinas}, \bibinfo{author}{Y.~Rath}, \bibinfo{author}{S.~Norimoto}, \bibinfo{author}{W.~Xie}, \bibinfo{author}{P.~See}, \bibinfo{author}{J.~P. Griffiths}, \bibinfo{author}{C.~Chen}, \bibinfo{author}{D.~A. Ritchie}, \bibinfo{author}{M.~Kataoka}, \bibinfo{author}{A.~Rossi}, \bibinfo{author}{I.~Rungger}, \bibinfo{journal}{Applied Physics Letters} \bibinfo{volume}{125} (\bibinfo{year}{2024}) \bibinfo{pages}{124001}.
\bibitem[{T\'oth et~al.(1999)T\'oth, Orlov, Amlani, Lent, Bernstein, and Snider}]{PhysRevB.60.16906}
\bibinfo{author}{G.~T\'oth}, \bibinfo{author}{A.~O. Orlov}, \bibinfo{author}{I.~Amlani}, \bibinfo{author}{C.~S. Lent}, \bibinfo{author}{G.~H. Bernstein}, \bibinfo{author}{G.~L. Snider}, \bibinfo{journal}{Phys. Rev. B} \bibinfo{volume}{60} (\bibinfo{year}{1999}) \bibinfo{pages}{16906--16912}.
\bibitem[{Fujisawa et~al.(2006)Fujisawa, Hayashi, and Sasaki}]{Fujisawa_2006}
\bibinfo{author}{T.~Fujisawa}, \bibinfo{author}{T.~Hayashi}, \bibinfo{author}{S.~Sasaki}, \bibinfo{journal}{Reports on Progress in Physics} \bibinfo{volume}{69} (\bibinfo{year}{2006}) \bibinfo{pages}{759}.
\bibitem[{Bäuerle et~al.(2018)Bäuerle, Christian~Glattli, Meunier, Portier, Roche, Roulleau, Takada, and Waintal}]{Bäuerle_2018}
\bibinfo{author}{C.~Bäuerle}, \bibinfo{author}{D.~Christian~Glattli}, \bibinfo{author}{T.~Meunier}, \bibinfo{author}{F.~Portier}, \bibinfo{author}{P.~Roche}, \bibinfo{author}{P.~Roulleau}, \bibinfo{author}{S.~Takada}, \bibinfo{author}{X.~Waintal}, \bibinfo{journal}{Reports on Progress in Physics} \bibinfo{volume}{81} (\bibinfo{year}{2018}) \bibinfo{pages}{056503}.
\bibitem[{Estrada Salda\~na et~al.(2018)Estrada Salda\~na, Vekris, Steffensen, \ifmmode~\check{Z}\else \v{Z}\fi{}itko, Krogstrup, Paaske, Grove-Rasmussen, and Nyg\aa{}rd}]{PhysRevLett.121.257701}
\bibinfo{author}{J.~C. Estrada Salda\~na}, \bibinfo{author}{A.~Vekris}, \bibinfo{author}{G.~Steffensen}, \bibinfo{author}{R.~\ifmmode~\check{Z}\else \v{Z}\fi{}itko}, \bibinfo{author}{P.~Krogstrup}, \bibinfo{author}{J.~Paaske}, \bibinfo{author}{K.~Grove-Rasmussen}, \bibinfo{author}{J.~Nyg\aa{}rd}, \bibinfo{journal}{Phys. Rev. Lett.} \bibinfo{volume}{121} (\bibinfo{year}{2018}) \bibinfo{pages}{257701}.
\bibitem[{Shang et~al.(2013)Shang, Li, Cao, Xiao, Tu, Jiang, Guo, and Guo}]{Shang2013}
\bibinfo{author}{R.~Shang}, \bibinfo{author}{H.-O. Li}, \bibinfo{author}{G.~Cao}, \bibinfo{author}{M.~Xiao}, \bibinfo{author}{T.~Tu}, \bibinfo{author}{H.~Jiang}, \bibinfo{author}{G.-C. Guo}, \bibinfo{author}{G.-P. Guo}, \bibinfo{journal}{Applied Physics Letters} \bibinfo{volume}{103} (\bibinfo{year}{2013}) \bibinfo{pages}{162109}.
\bibitem[{Banszerus et~al.(2020)Banszerus, M{\"o}ller, Icking, Watanabe, Taniguchi, Volk, and Stampfer}]{Banszerus2020}
\bibinfo{author}{L.~Banszerus}, \bibinfo{author}{S.~M{\"o}ller}, \bibinfo{author}{E.~Icking}, \bibinfo{author}{K.~Watanabe}, \bibinfo{author}{T.~Taniguchi}, \bibinfo{author}{C.~Volk}, \bibinfo{author}{C.~Stampfer}, \bibinfo{journal}{Nano Letters} \bibinfo{volume}{20} (\bibinfo{year}{2020}) \bibinfo{pages}{2005--2011}. \bibinfo{note}{PMID: 32083885}.
\bibitem[{Yamahata et~al.(2023)Yamahata, Johnson, and Fujiwara}]{PhysRevApplied.20.044043}
\bibinfo{author}{G.~Yamahata}, \bibinfo{author}{N.~Johnson}, \bibinfo{author}{A.~Fujiwara}, \bibinfo{journal}{Phys. Rev. Appl.} \bibinfo{volume}{20} (\bibinfo{year}{2023}) \bibinfo{pages}{044043}.
\bibitem[{Limbach(2002)}]{Limbach2002}
\bibinfo{author}{B.~Limbach}, \bibinfo{title}{Strong tunneling in metallic double island structure}, Ph.D. thesis, University Karlsruhe (TH), \bibinfo{address}{Karlsruhe}, \bibinfo{year}{2002}.
\bibitem[{Wallisser et~al.(2002)Wallisser, Limbach, vom Stein, Sch\"afer, Theis, G\"oppert, and Grabert}]{Wallisser2002}
\bibinfo{author}{C.~Wallisser}, \bibinfo{author}{B.~Limbach}, \bibinfo{author}{P.~vom Stein}, \bibinfo{author}{R.~Sch\"afer}, \bibinfo{author}{C.~Theis}, \bibinfo{author}{G.~G\"oppert}, \bibinfo{author}{H.~Grabert}, \bibinfo{journal}{Phys. Rev. B} \bibinfo{volume}{66} (\bibinfo{year}{2002}) \bibinfo{pages}{125314}.
\bibitem[{Theis(2004)}]{Christoph2004}
\bibinfo{author}{C.~Theis}, \bibinfo{title}{Conductance of Single Electron Devices from Imaginary -Time Path Integrals}, \bibinfo{type}{{PhD} dissertation}, Freiburg (Breisgau), University, Dissertation, available as pdf. under https://freidok.uni-freiburg.de/fedora/objects/freidok:1328/datastreams/FILE1/content, \bibinfo{year}{2004}.
\bibitem[{van~der Wiel et~al.(2002)van~der Wiel, De~Franceschi, Elzerman, Fujisawa, Tarucha, and Kouwenhoven}]{Fujisawa2002}
\bibinfo{author}{W.~G. van~der Wiel}, \bibinfo{author}{S.~De~Franceschi}, \bibinfo{author}{J.~M. Elzerman}, \bibinfo{author}{T.~Fujisawa}, \bibinfo{author}{S.~Tarucha}, \bibinfo{author}{L.~P. Kouwenhoven}, \bibinfo{journal}{Rev. Mod. Phys.} \bibinfo{volume}{75} (\bibinfo{year}{2002}) \bibinfo{pages}{1--22}.
\bibitem[{Kubo(1966)}]{Kubo_1966}
\bibinfo{author}{R.~Kubo}, \bibinfo{journal}{Reports on Progress in Physics} \bibinfo{volume}{29} (\bibinfo{year}{1966}) \bibinfo{pages}{255}.
\bibitem[{G\"oppert et~al.(1999)G\"oppert, Grabert, and Beck}]{Goppert1999}
\bibinfo{author}{G.~G\"oppert}, \bibinfo{author}{H.~Grabert}, \bibinfo{author}{C.~Beck}, \bibinfo{journal}{Europhys. Lett.} \bibinfo{volume}{45} (\bibinfo{year}{1999}) \bibinfo{pages}{249--255}.
\bibitem[{Negele and Orland(1987)}]{Negele1987}
\bibinfo{author}{J.~W. Negele}, \bibinfo{author}{H.~Orland}, \bibinfo{title}{Quantum Many-Particle Systems (Frontier in Physics)}, \bibinfo{publisher}{Addison{Wesley}}, \bibinfo{year}{1987}.
\bibitem[{G\"oppert et~al.(2000)G\"oppert, H\"upper, and Grabert}]{Goppert2000}
\bibinfo{author}{G.~G\"oppert}, \bibinfo{author}{B.~H\"upper}, \bibinfo{author}{H.~Grabert}, \bibinfo{journal}{Phys. Rev. B} \bibinfo{volume}{62} (\bibinfo{year}{2000}) \bibinfo{pages}{9955--9958}.
\bibitem[{Harata and Srivilai(2022)}]{Harata_2022}
\bibinfo{author}{P.~Harata}, \bibinfo{author}{P.~Srivilai}, \bibinfo{journal}{Journal of Statistical Mechanics: Theory and Experiment} \bibinfo{volume}{2022} (\bibinfo{year}{2022}) \bibinfo{pages}{013101}.
\bibitem[{G\"oppert and Grabert(1998)}]{PhysRevB.58.R10155}
\bibinfo{author}{G.~G\"oppert}, \bibinfo{author}{H.~Grabert}, \bibinfo{journal}{Phys. Rev. B} \bibinfo{volume}{58} (\bibinfo{year}{1998}) \bibinfo{pages}{R10155--R10158}.
\bibitem[{Srivilai(2012)}]{Prathan2012}
\bibinfo{author}{P.~Srivilai}, \bibinfo{title}{Quantum Monte Carlo Study of the Metallic Single Electron Pump}, \bibinfo{type}{{PhD} dissertation}, Freiburg (Breisgau), University, Dissertation, \bibinfo{year}{2012}.
\bibitem[{Joyez et~al.(1997)Joyez, Bouchiat, Esteve, Urbina, and Devoret}]{Joyez1997}
\bibinfo{author}{P.~Joyez}, \bibinfo{author}{V.~Bouchiat}, \bibinfo{author}{D.~Esteve}, \bibinfo{author}{C.~Urbina}, \bibinfo{author}{M.~H. Devoret}, \bibinfo{journal}{Phys. Rev. Lett.} \bibinfo{volume}{79} (\bibinfo{year}{1997}) \bibinfo{pages}{1349--1352}.
\bibitem[{Hansen(1987)}]{hansen87}
\bibinfo{author}{P.~C. Hansen}, \bibinfo{journal}{BIT} \bibinfo{volume}{27} (\bibinfo{year}{1987}) \bibinfo{pages}{534--553}.
\bibitem[{Louis(1989)}]{louis89}
\bibinfo{author}{A.~K. Louis}, \bibinfo{title}{Inverse und schlecht gestellte Probleme}, Teubner Studienb\"{u}cher: Mathematik, \bibinfo{publisher}{B.G.Teubner}, \bibinfo{address}{Stuttgart}, \bibinfo{year}{1989}.
\bibitem[{Press et~al.(1992)Press, Teukolsky, Vetterling, and Flannery}]{press92}
\bibinfo{author}{W.~H. Press}, \bibinfo{author}{S.~A. Teukolsky}, \bibinfo{author}{W.~T. Vetterling}, \bibinfo{author}{B.~P. Flannery}, \bibinfo{title}{Numerical Recipes in C - The Art of Scientific Computing}, \bibinfo{publisher}{Cambridge University Press}, \bibinfo{address}{Cambridge}, \bibinfo{edition}{2nd corrected edition} edition, \bibinfo{year}{1992}.
\bibitem[{Ceperley(1995)}]{Ceperley1995}
\bibinfo{author}{D.~M. Ceperley}, \bibinfo{journal}{Rev. Mod. Phys.} \bibinfo{volume}{67} (\bibinfo{year}{1995}) \bibinfo{pages}{279--355}.
\bibitem[{Troyer and Wiese(2005)}]{Troyer2005}
\bibinfo{author}{M.~Troyer}, \bibinfo{author}{U.-J. Wiese}, \bibinfo{journal}{Phys. Rev. Lett.} \bibinfo{volume}{94} (\bibinfo{year}{2005}) \bibinfo{pages}{170201}.
\bibitem[{Sergei~L. and Werner(2006)}]{Lukyanov2006}
\bibinfo{author}{L.~Sergei~L.}, \bibinfo{author}{P.~Werner}, \bibinfo{journal}{Journal of Statistical Mechanics: Theory and Experiment} \bibinfo{volume}{2006} (\bibinfo{year}{2006}) \bibinfo{pages}{11002}.

\end{thebibliography}

\end{document}